\documentclass[conference]{IEEEtran}
\usepackage{cite}
\usepackage{amsmath,amssymb,amsfonts}
\usepackage{graphicx}
\usepackage{textcomp}
\usepackage{xcolor}

\newcommand\copyrighttext{%
  \footnotesize \textcopyright 2020 IEEE. Personal use of this material is permitted.  Permission from IEEE must be obtained for all other uses, in any current or future media, including reprinting/republishing this material for advertising or promotional purposes, creating new collective works, for resale or redistribution to servers or lists, or reuse of any copyrighted component of this work in other works.}
\newcommand\copyrightnotice{%
\begin{tikzpicture}[remember picture,overlay]
\node[anchor=south,yshift=10pt] at (current page.south) {\fbox{\parbox{\dimexpr\textwidth-\fboxsep-\fboxrule\relax}{\copyrighttext}}};
\end{tikzpicture}%
}

\usepackage{environ}
\usepackage[linesnumbered,ruled,vlined]{algorithm2e}
\SetVlineSkip{1pt}
\NewEnviron{myblock}[1]{
    \SetKwBlock{mybegin}{#1}{}
    \mybegin{\BODY}
}
\SetKwSty{textnormal}

\usepackage{etoolbox}

\usepackage[utf8]{inputenc}

\usepackage{enumitem}

\ifCLASSOPTIONcompsoc 
    \usepackage[caption=false,font=normalsize,labelfont=sf,textfont=sf]{subfig}
\else
    \usepackage[caption=false,font=footnotesize]{subfig}
\fi

 \colorlet{mylinkcolor}{red}
 \colorlet{mycitecolor}{green}
 \colorlet{myurlcolor}{magenta}
 \usepackage[
     colorlinks  = true,
     linkcolor   = mylinkcolor!80!black,
     citecolor   = mycitecolor!60!black,
     urlcolor    = myurlcolor!70!black,
 ]{hyperref}
 \usepackage{url}

\usepackage[capitalize, nameinlink]{cleveref}
\crefname{section}{Sec.}{Secs.}
\crefname{appendix}{App.}{Apps.}
\crefname{figure}{Fig.}{Figs.}
\crefname{algorithm}{Alg.}{Algs.}

\usepackage{float} 

\DeclareFontFamily{U}{jkpmia}{}
\DeclareFontShape{U}{jkpmia}{m}{it}{<->s*jkpmia}{}
\DeclareFontShape{U}{jkpmia}{bx}{it}{<->s*jkpbmia}{}
\DeclareMathAlphabet{\mathfrak}{U}{jkpmia}{m}{it}
\SetMathAlphabet{\mathfrak}{bold}{U}{jkpmia}{bx}{it}
\newcommand{\p}{\mathfrak{p}}

\usepackage{mathtools}

\usepackage{microtype} 

\usepackage{multirow,array} 

\usepackage{ntheorem} 
\theorembodyfont{\upshape}
\newtheorem{definition}{Definition}
    \crefname{definition}{Def.}{Defs.}
    \crefname{myalgorithm}{Alg.}{Algs.}
\newtheorem{example}{Example}
    \crefname{example}{Ex.}{Exs.}

\usepackage[ntheorem]{mdframed}
\theoremstyle{empty}
\newmdtheoremenv[linewidth=3pt, leftmargin=1pt, linecolor=black!20!white, rightline=false, topline=false, bottomline=false, skipabove=9pt, skipbelow=20pt]{blank}{Theorem}

\newenvironment{offset}
{
    \begin{blank}
    \vspace{-3pt}
}%
{
    \vspace{-3pt}
    \end{blank}
    \smallskip
}

\usepackage{tikz}

\usepackage{dsfont} 

\newcommand\mydots{\makebox[1em][c]{.\hfil.\hfil.}\thinspace}
\newcommand\mymdots{\makebox[1em][c]{.\hfil.\hfil.}}

\begin{document}

\title{Formal Game Grammar and Equivalence
}

\author{\IEEEauthorblockN{Paul Riggins}
\IEEEauthorblockA{\textit{Berkeley Center for Theoretical Physics}\\
\textit{University of California}\\
Berkeley, CA, USA\\
priggins@berkeley.edu}
\and
\IEEEauthorblockN{David McPherson}
\IEEEauthorblockA{\textit{Department of Electrical Engineering and Computer Sciences}\\
\textit{University of California}\\
Berkeley, CA, USA\\
david.mcpherson@eecs.berkeley.edu}%
}

\IEEEoverridecommandlockouts
\IEEEpubid{\begin{minipage}{\textwidth}\ \\[12pt]
978-1-7281-4533-4/20/\$31.00 \copyright 2020 IEEE
\end{minipage}}

\maketitle
\copyrightnotice
\begin{abstract}
    We develop methods to formally describe and compare games, in order to probe questions of game structure and design%
, and as a stepping stone to predicting player behavior from design patterns.
    We define a grammar-like formalism to describe finite discrete games without hidden information, allowing for randomness, and mixed sequential and simultaneous play.
    We make minimal assumptions about the form or content of game rules or user interface.
    The associated game trees resemble hybrid extensive- and strategic-form games, in the game theory sense. 
    By transforming and comparing game trees, we develop equivalence relations on the space of game systems, which equate games that give players the same meaningful agency.
    We bring these together to suggest a method to measure distance between games, insensitive to cosmetic variations in the game logic descriptions.
\end{abstract}
\begin{IEEEkeywords}
    game grammar, game representation, game tree, equivalence relations, similarity measures, mathematical ludology
\end{IEEEkeywords}
\vspace{-6pt}

\section{Introduction}
\label{sec:Introduction}

Games are notable among artistic media in that the interactive systems underlying them can often be precisely defined.
This has enabled great mathematical progress in understanding game-centric decision-making processes, through efforts in game theory and artificial intelligence (AI) (e.g., \cite{yannakakis_artificial_2018}), but this progress has been largely isolated from other, ``softer'' subfields of game studies \cite{melcer_games_2015,melcer_toward_2017}.
In particular, little mathematical attention has been given to questions of game design, even while the methods and vocabulary used by designers has become increasingly sophisticated and systematic (e.g., \cite{tekinbas_rules_2003,koster_theory_2005,adams_game_2012,schell_art_2019,engelstein_building_2019}).
The precise definability of games, however, could also be used to formally explore questions of interest to ludologers and designers:
How much does a game mechanic matter for overall gameplay?
What's the best user interface to reflect the underlying rules?
Can we predict behavior in one game based on behavior in a similar game?

In \cite{riggins_tools_2019}, we proposed the study of mathematical ludology, aiming to bring mathematical attention to questions like these by formally exploring the space of games and their properties.
This is, in some sense, a mathematically formal continuation of ``game grammar'' efforts begun in the game design community to atomize and interrogate game designs (e.g., \cite{cousins_elementary_2004,koster_grammar_2005,adams_game_2012,stephane_game_2006}).
In this paper%
\footnotemark 
we especially develop and explore formal notions of equivalence and similarity between games.

The ability to compare games for equivalence or similarity could be useful for studying relationships, building taxonomies, transferring behavioral learning (perhaps AI learning) from one game to another, or maybe even developing approximate game theoretic solutions for similar games.
A key challenge, however, lies in the many ways the same game can be described, even with the same formalism.
For instance, we are aware of one other complementary work-in-progress aiming to measure distance between games: the Digital Ludeme Project, advancing archaeoludology \cite{browne_modern_2018}.
The proposed method, using edit distance between rule trees, is powerful but sensitive to cosmetic differences in the game rules: two different descriptions could describe identical games, yet have nonzero distance.
The senses of equivalence and similarity we propose here are specifically designed to avoid this aesthetic sensitivity.

The key \textbf{contributions} of this paper are
\begin{enumerate}
    \item 
        We develop a grammar-like formalism to describe finite, discrete games (\cref{sec:UnderlyingGameSystem}), more flexible than game theoretic methods and more tailored for abstract structural analysis than general gameplaying (GGP) approaches.
    \item 
After constructing game trees (\cref{sec:GameTreesAndAutomata}), we develop equivalence relations on the spaces of game trees and game systems (\cref{sec:AgencyEquivalence}).
    \item 
We bring these together to suggest a way to measure game similarity---insensitive to cosmetic variations in rule descriptions---by sampling the game state spaces and checking for equivalence of partial game trees (\cref{sec:Discussion}).
\end{enumerate}

\footnotetext{Most of the content in this paper first appeared in the preprint \cite{riggins_tools_2019}.}

\vspace{-1pt}
\section{Underlying Game Systems}
\label{sec:UnderlyingGameSystem}

\begin{figure}
    \centering
    \includegraphics[scale=0.6]{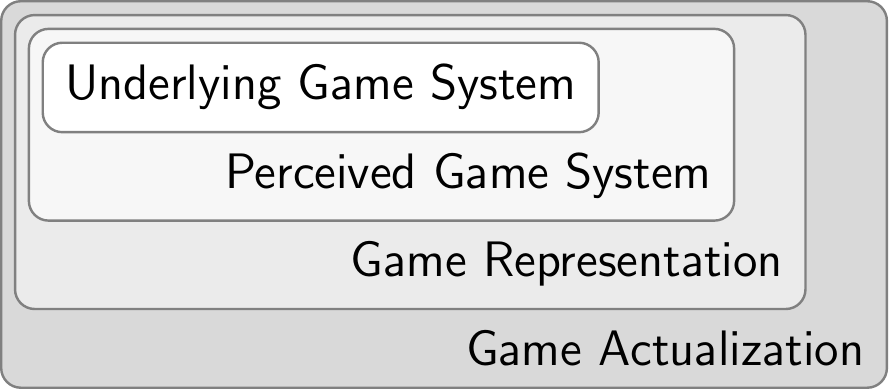}
    \caption{A game description hierarchy (see \cite{riggins_tools_2019}).
        Underlying Game Systems contain the essential logic of a game.
        (These are the focus of this paper.)
        Perceived Game Systems add an information layer, specifying what information about the game is given to each player, and how this interacts with the game logic.
        Game Representations add a user interface specification, how the game is represented to players, e.g., visually.
        Game Actualizations are the real-world realizations, like a physical chess set or software.
        Player preferences, skill levels, etc., are relegated to Player Models, outside these game descriptions.
    } 
    \label{fig:GameDescriptionHierarchy}
    \vspace{-10pt}
\end{figure}

The present paper will focus on the essential logic of games, what we will call the \emph{(underlying) game system}.%
\footnote{Comparable to the ``constituative [\emph{sic}] rules'' of \cite{tekinbas_rules_2003}.}
This is the base of a game description hierarchy described in \cite{riggins_tools_2019} (see \cref{fig:GameDescriptionHierarchy}), and in particular does not specify the information available to (or hidden from) players, or the form of the user interface---these are naturally quite important, but also add many complexities.
An underlying game system provides the game logic as an omniscient referee might see it---as a (nondeterministic) game of perfect information---which is interesting in its own right and a useful foundation to build on.
In contrast to game theoretic approaches, we do not include player preferences or payoffs (i.e., how individual players value different outcomes) in our game descriptions.
Thus we draw a formal separation between game and players, adhering more closely to the colloquial understanding of a game.

Specifically, we will focus on games that are discrete in time and space and that have a finite state space; \cref{def:UnderlyingGameSystem} can describe the game system of any such game.
This captures the majority of board and card games, and many video games.
\cref{def:UnderlyingGameSystem} is not the only way to describe a game system, but it captures those minimal elements essential to the logic of a game, while also making formally clear what agency each player has versus what is outside their control.

In particular, \cref{def:UnderlyingGameSystem} provides: the players ($\mathcal{P}$), the game state space ($\mathbb{S}$, factorized via $\mathcal{T}$), initial game states ($\mathcal{S}_0$), decisions available to players ($\mathcal{D}$), when those decisions are legal ($L$), the possible consequences of those decisions ($C$), how the game state is changed as a result ($\mathcal{A}$), and possible game outcomes ($\mathcal{O}, \Omega$).
It allows for mixed sequential and simultaneous play, deterministic or nondeterministic, and makes no additional assumptions about the content of games, not even presuming the existence of boards, pieces, or alternating turns.

\begin{definition}
    \label{def:UnderlyingGameSystem}
    An \emph{(underlying) game system} $\mathcal{G}$ with $n$ players is a $9$-tuple $\mathcal{G} = \langle \mathcal{P}, \mathcal{T}, \mathcal{S}_0, \mathcal{D}, \mathcal{A}, C, L, \mathcal{O}, \Omega \rangle$, where:
  \begin{itemize}
    \item $\mathcal{P} = (1, \ldots, n)$ is a list of \emph{players}, agents which may make decisions in the game.
    \item $\mathcal{T} = (T_1, \ldots, T_m)$ is a finite list of finite sets, called \emph{substate tracks}. The set of \emph{game states} $\mathbb{S}$ is given by $\mathbb{S} = T_1 \times \cdots \times T_m.$
    \item $\mathcal{S}_0 \subset \mathbb{S}$ is a set of \emph{initial conditions}.
    \item $\mathcal{D}$ is a set of \emph{decisions}, the choices which players may make in order to influence (but not directly change) the game state. This is extended with the \emph{null decision} $0 \notin \mathcal{D}$ to form $\mathcal{D}_0 \equiv \mathcal{D}\cup\{0\}$.
    \item $\mathcal{A}$ is a set of \emph{actions} $a: \mathbb{S} \to \mathbb{S}$, which can directly modify the game state. 
    \item $C$ is a \emph{consequence function} $C(d_0^n, s)$ which takes a decision tuple $d_0^n\in\mathcal{D}_0^n$ (i.e., one possibly null decision per player) and state $s\in\mathbb{S}$ and returns a nonempty set of \emph{consequences}: a set of pairs $(\p_a, a)$, where $\p_a \in (0,1]$ is a non-zero probability and $a$ is an action or product of actions. The sum of probabilities in the set must equal 1.
        These are the consequences of decisions, which may be outside any one player's control.
    \item $L: \mathcal{P} \times \mathcal{D}\to 2^{\mathbb{S}}$ is a \emph{legality function}, which returns a (possibly empty) subset of $\mathbb{S}$ for each player $p\in\mathcal{P}$ and decision $d\in \mathcal{D}$, reflecting when that player can make that decision. 
        The \emph{legal set of decisions for player $p$ at state $s$} is the set $L_p(s) \equiv \{ d\in\mathcal{D}: s \in L(p,d) \}$.
        A decision $d \in L_p(s)$ is \emph{legal} for $p$ at $s$, and \emph{illegal} otherwise.
    \item $\mathcal{O}$ is the set of \emph{outcomes} that can result from the game.
    \item $\Omega$ is an \emph{outcome function} $\Omega: \mathcal{S}_\text{ter}\to\mathcal{O}$, where $\mathcal{S}_\text{ter} \equiv \{ s\in\mathbb{S}: L_p(s) = \varnothing \text{ for all }p\in\mathcal{P}\}$ is the set of \emph{terminal game states}.
        Intuitively, $\mathcal{S}_\text{ter}$ is the set of game states at which no legal decisions can be made, so the game ends and the result is computed by $\Omega$.
  \end{itemize}
\end{definition}

The separation of decisions (player choices) from actions (changes to the game state) via consequence functions provides a formal separation between what individual players can and cannot control.
Consequence functions are necessary to handle random chance and simultaneous play.
In both cases, each player can influence play by their chosen decision, but the ultimate effect on the game state (the consequent action) is determined probabilistically and/or after considering the simultaneous decisions of other players (see \cref{def:GameplayAlgorithm,ex:TicTacToe}).
In sequential, deterministic games, it is possible to have a one-to-one mapping between decisions and actions---individual players can directly determine game state changes---and so consequence functions are conceptually superfluous.

\SetAlgoHangIndent{4pt}
\SetInd{3pt}{5pt}
\makeatletter
\patchcmd{\@algocf@start}
  {-1.5em}
  {-5pt}
  {}{}
\makeatother

The following algorithm can be used (by players) to play any game with an underlying game system described by \cref{def:UnderlyingGameSystem}:
\vspace{-16pt}
\begin{algorithm}
            All players agree on some $s_0\in\mathcal{S}_0$. Let the current state be $s' = s_0$.
            \smallskip

            \begin{myblock}{\textbf{While} $s'$ is not a terminal state ($s' \notin \mathcal{S}_\text{ter}$), repeat:}
            \smallskip
                Each player $p$ selects one decision from their respective legal set $L_p(s')$ at the current state. If $L_p(s') = \varnothing$ for a player $p$, then $p$ is assigned the null decision: $d_p = 0$.

            \smallskip

                Compute the set of consequences from the decision tuple and the current state: $c = C((d_1, \ldots, d_n), s') =$ $ \{(\p_1, a_1), \ldots, (\p_m, a_m)\}$.
                    Randomly pick a consequent action from $c$, where $a_j$ is chosen with probability $\p_j$.
            \smallskip

                Compute the new game state $s'' = a_j s'$. Let $s' = s''$.
        \end{myblock}
            \smallskip
        The game is over ($s' \in \mathcal{S}_\text{ter}$). Compute the outcome $\Omega(s')$.

    \caption{Gameplay Algorithm}
    \label{def:GameplayAlgorithm}
\end{algorithm}
\vspace{-10pt}

If this algorithm can always be faithfully executed for a game system, that system is \emph{complete} \cite{riggins_tools_2019}, and its game tree(s) can be built (see \cref{sec:GameTreesAndAutomata}).

\begin{figure}
    \setlength{\abovecaptionskip}{-7pt}
    \centering
    \vspace{-4pt}
    \includegraphics[width=\columnwidth]{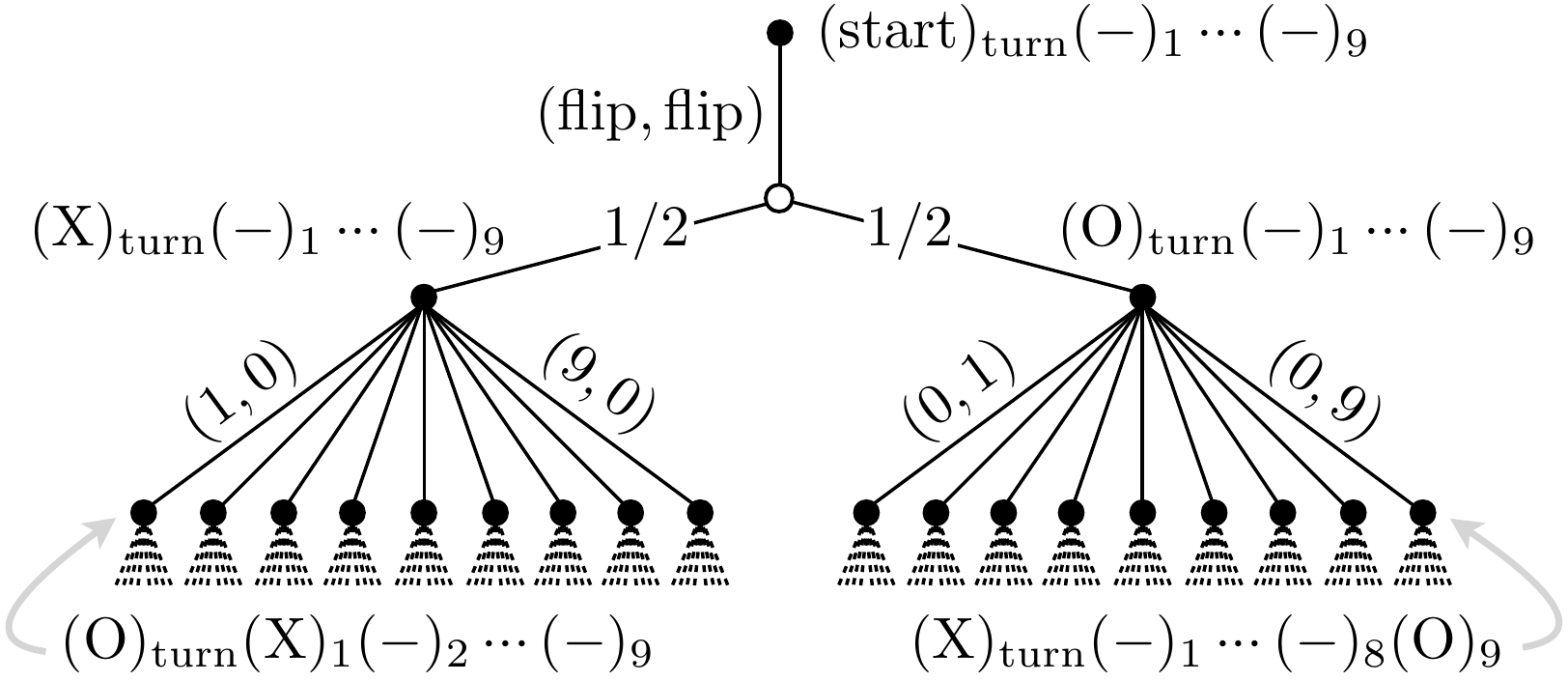}
    \caption{The first couple generations of the game tree for the tic-tac-toe system in \cref{ex:TicTacToe}, with only a few state nodes and decision edges labeled, for brevity.
        Solid nodes are state nodes, while the open node is a chance node.
        The root state node has the initial state from $\mathcal{S}_0$.
        The first decision edge with tuple $(\text{flip},\text{flip})$ leads to a chance node, with outgoing chance edges both with probability $1/2$.
        Subsequent decisions claim numbers for a player and toggle the turn track, e.g., $(1,0)$ claims 1 for X; from the resulting state $(\text{O})_\text{turn}(\text{X})_1(-)_2\cdots(-)_9$, there are 8 legal decision tuples $(0,2), \mydots, (0,9)$ as the game tree continues, which would each claim a number for player O.
    }
    \label{fig:CoinFlipGameTree}
    \vspace{-9pt}
\end{figure}

Here is an example of a tic-tac-toe variant, with basic notation described along the way.
Its game tree is illustrated in \cref{fig:CoinFlipGameTree}.
More complex games (or even this one) can benefit from richer notation in order to write them more compactly.
Some notational suggestions and examples can be found in \cite{riggins_tools_2019}; developing notation is not the purpose of the present paper.

\begin{figure}[b]
    \vspace{-13pt}
  \begin{minipage}[c]{0.4\linewidth}
    \centering
    \small
    \begin{tabular}{|r|r|r|}\hline
        $2$ & $7$  & $6$ \\\hline
        $9$  & $5$  & $1$ \\\hline
        $4$ & $3$ & $8$  \\\hline
    \end{tabular}
  \end{minipage}%
  \begin{minipage}[c]{0.5\linewidth}
    \caption{``Magic square'' correspondence between tic-tac-toe and 3-to-15.} 
    \vspace{6pt}
    \label{fig:MagicSquare}
  \end{minipage}
  \vspace{-2pt}
\end{figure}

\begin{example}[Tic-Tac-Toe with random start]
    \label{ex:TicTacToe}
    Two players play tic-tac-toe (or ``3-to-15''), randomly choosing who starts.

\emph{(States.)} We write $(v)_t$ to express the subset of $\mathbb{S}$ where track $T_t$ takes value $v$; product and sum notation on these subsets denote intersection and union, respectively.

\emph{(Actions.)} By $a: S\mapsto (v_1)_1\cdots(v_k)_k$, we mean that for any state in the subset $S \subset \mathbb{S}$, the action $a \in \mathcal{A}$ changes the value of track $T_1$ to $v_1$, and so on to track $T_k$, acting as the identity on any tracks not appearing on the right-hand side.
It also acts as the identity on any state $s\notin S$.
E.g., if $a: \mathbb{S}\mapsto (1)_a$ and we have some state $s = (2)_a(3)_b$, then $a\cdot s = (1)_a(3)_b$.

\emph{(Outcome functions.)}
We take $\Omega: S\mapsto \omega$ to mean $\Omega(z) = \omega$ for all terminal states $z \in S \subset \mathbb{S}$.

We'll break up the game description for ease of exposition.
There are two players and 10 available decisions: one for each space, plus the coin flip. 
There are 10 tracks to record the turn and spaces, and the initial condition is an empty board.
\vspace{-2pt}
\begin{offset}
    $\mathcal{P} = \{ \text{X}, \text{O} \}$, \quad  $\mathcal{D} = \{ 1, \mydots, 9, \text{flip} \}$ \\[2pt]
    $\makebox[7mm]{$\mathcal{T}:$\hfill} \makebox[7mm]{$T_\text{turn}$\hfill} = \{ \text{start}, \text{X}, \text{O} \},$ \\ 
    $\makebox[7mm]{} \makebox[7mm]{$T_{i}$\hfill} = \{ -, \text{X}, \text{O} \}\ \text{ for }\ i\in\{ 1, \mydots, 9 \}$ \\[2pt]
    $\mathcal{S}_0 = (\text{start})_\text{turn} (-)_{1} \cdots (-)_{9}$
\end{offset}
From the initial condition, both players can only legally ($L$) choose the decision ``flip''.
As a consequence ($C$) of this joint decision, the value of track $T_\text{turn}$ becomes X or O with a 50-50 chance, via the action ``X first'' or ``O first''.
\begin{offset}
    $L(p, \text{flip}) = (\text{start})_\text{turn}, \quad p \in \mathcal{P}$ \\[2pt]
    $C((\text{flip}, \text{flip}), \mathbb{S}) = \{ (1/2, \text{X first}), (1/2, \text{O first}) \}$ \\[2pt]
    $\mathcal{A} \supset \{ \text{X first, O first} \},\quad p\text{ first}: (\text{start})_\text{turn} \mapsto (p)_\text{turn}$
\end{offset}
Players then alternate (enforced by $L(p,i)$ and the ``turn'' track), picking unclaimed (``$-$'') spaces via the decisions $1,\mydots,9 \in \mathcal{D}$ until either no more spaces are available or someone wins. 
\begin{offset}
    $L(p, i) = (p)_\text{turn} (-)_i\ \overline{E}, \qquad  i\in\{ 1, \mydots, 9 \} $ \\[2pt]
    $C((i, 0), \mathbb{S}) = (1, \text{X}_{i} \cdot \text{next}),\ \ C((0, i), \mathbb{S}) = (1, \text{O}_{i} \cdot \text{next}) $ \\[2pt]
    $\mathcal{A} \supset \{ \text{X}_{1}, \text{O}_{1}, \mydots, \text{X}_9, \text{O}_9 \},\quad p_i: (-)_i \mapsto (p)_i $ \\[2pt]
    $\mathcal{A} \supset \{ \text{next} \},\ \ \text{next}: (\text{X})_\text{turn} \mapsto (\text{O})_\text{turn},\ (\text{O})_\text{turn} \mapsto (\text{X})_\text{turn}$ 
    \vspace{-9pt}
\end{offset}
The game then ends in victory or draw, respectively.
Note we've defined an auxiliary set $E$ to compactly capture winning ending states.
(The complement $\overline{E}$ appears in $L(p,i)$, so play can only legally proceed if a victory state has not been reached.)
\begin{offset}
    $\displaystyle E = E_\text{X} \,\cup\, E_{O}, \hfill E_p = \hspace{-4pt} \sum_{\substack{i,j,k \in \{1,\mydots,9\}\\i+j+k=15}} \hspace{-4pt} (p)_i(p)_j(p)_k $ \hspace{-10pt} \\[-12pt]
    $\mathcal{O} = \{ \text{X wins}, \text{O wins}, \text{draw} \}$ \\[4pt]
    $\Omega: E_p \mapsto p\text{ wins}, \quad \text{otherwise} \mapsto \text{draw}$
\end{offset}

In tic-tac-toe, the ending states correspond to three-in-a-row board states.
In 3-to-15, these correspond to integer triples that sum to fifteen.
These games may be played with different user interfaces, but they share the same essential logic (see \cref{fig:MagicSquare}), and thus the same underlying game system.

\end{example}

\subsection{Why not use an existing game description formalism?}
\label{sec:ExistingDescriptions}

Game theoretic descriptions are too limited in the games they can practically express.
Strategic- and extensive-form games respectively describe simultaneous and sequential games well, but lose important nuance when trying to mix the two \cite{cooper_communication_1992,salles_beyond_2008}. Additionally, complex game descriptions are intractable with game theoretic formal descriptions.
Generally, either the full game can be written explicitly as a strategic, extensive, or combinatorial game (all intractable, e.g., for chess), or else game theorists rely on ad hoc natural language description and reader familiarity to communicate the rules before proceeding to analysis (e.g., \cite{beck_combinatorial_2008}). 
We want a way to formally, and tractably, describe game rules even for complex games; a grammar-like formalism is better suited to this task.

We also desire a total conceptual separation between game and player descriptions; game theory does not make this separation.
This separation is also why we have included randomness in the game description itself---in contrast to typical game theory or GGP formalisms, which invoke an extra fictional player who behaves randomly \cite{rasmusen_games_2007,piette_ludii_2019,thielscher_general_2010}.

GGP descriptions like GDL \cite{love_general_2006,thielscher_general_2010,thielscher_gdl-iii:_2017}, RBG \cite{kowalski_regular_2019}, or Ludii \cite{piette_ludii_2019} are designed especially for efficient software implementation and AI methods, often with particular classes of games in mind.
Ludii and RBG have compact notation, but require the construction of a game board (i.e., visual user interface) integrated with the rules, distinguishing them most naturally as game representations (see \cref{fig:GameDescriptionHierarchy}) suited to traditional board games; we desire a lower-level description to study and compare game rules, which makes fewer assumptions.
GDL is very generic, like we might prefer, but can be intractably verbose, with an opaque state space.
Our \cref{def:UnderlyingGameSystem} bears some formal similarity to GDL-II without hidden information, and is just as expressive, but offers a simpler specification of state space, a different treatment of randomness, and we permit a more compact and extensible notation.

\section{Game Trees}
\label{sec:GameTreesAndAutomata}

A complete game system from \cref{def:UnderlyingGameSystem} can be used to generate a (possibly infinite) game tree or a finite (possibly nondeterministic) game automaton.
These are useful for visualizing and analyzing the game systems, as well as for making connection with existing work in game theory and AI.
They are graphical representations of \cref{def:GameplayAlgorithm}: each playthrough from that algorithm identifies a path from an initial node to a terminal node in a tree or automaton.
We will focus on game trees in this paper.

\SetAlgoHangIndent{4pt}
\SetInd{3pt}{4pt}
\makeatletter
\patchcmd{\@algocf@start}
  {-5pt}
  {0pt}
  {}{}
\makeatother

\begin{definition}
    \label{def:GameSystemTree}
    The \emph{game trees} of a complete game system $\mathcal{G}$ is a set of game trees $\tau(\mathcal{G}) \equiv \{ \tau(\mathcal{G}|s_0): s_0\in\mathcal{S}_0 \}$, one for each initial condition. 
    Each game tree $\tau(\mathcal{G}|s_0)$ is built via \cref{alg:GameTree}.
\end{definition}

\begin{algorithm}
    Draw a root node, assigned the initial state $s_0$.
    \smallskip

    \begin{myblock}{\textbf{While} not all leaves have assigned outcomes, repeat:}
        \smallskip

    \begin{myblock}{\textbf{For} each leaf node $w$ in the current tree, with assigned state $s(w)$ but no assigned outcome, do the following:}
        \smallskip

        Let $s = s(w)$. 
        \textbf{If} $s \in \mathcal{S}_\text{ter}$, \textbf{then} assign the outcome $\Omega(s)$ to $w$, and stop for node $w$. \textbf{Else}, proceed:
        \smallskip

        Generate all legal decision tuples at state $s$ from the legal set $L_p(s)$ for each player: $D_0^n(s) \equiv \{ (d_1, \mymdots, d_n)\hspace{-2pt}:$ $d_p = 0 \text{ if } L_p(s) = \varnothing,\text{ else }d_p\in L_p(s) \}$.
        \smallskip

        \begin{myblock}{\textbf{For} each tuple $d_0^n \in D_0^n(s)$, do the following:}
            \smallskip
            
                Draw a child node $w'$ below $w$, with a directed edge from $w$ to $w'$.
            Assign $d_0^n$ to this edge.
            \smallskip

                Compute the set of consequences $c' = C(d_0^n,s)$. 
                \smallskip
                
                \textbf{If} $c' = \{ (1, a) \}$, \textbf{then} assign the state $a\cdot s$ to $w'$\hspace{-2pt}.
                \textbf{Else}:\hspace{-2pt}
                    \smallskip

                \textbf{For} each probability-action pair $(\p_i,a_i) \in c'$, draw a child node $w_i''$ with a directed edge from $w'$ to $w_i''$.
                    Assign $\p_i$ to this edge, and the state $a_i\cdot s$ to $w_i''$.
            \end{myblock}
    \end{myblock}
\end{myblock}
\caption{Game Tree Construction, for $\tau(\mathcal{G} | s_0)$}
\label{alg:GameTree}
  \end{algorithm}

Each tree resulting from \cref{alg:GameTree} has the following structure:
\emph{State nodes} have assigned states and outgoing \emph{decision edges}, which have assigned decision tuples.
These decision edges lead to either new state nodes, or to unlabeled \emph{chance nodes} which have outgoing \emph{chance edges} labeled with probabilities.
These chance edges lead to new state nodes.
State nodes may be further subdivided into \emph{single-player nodes}, in which only one player has legal decisions available (the node \emph{belongs} to that player); \emph{multiplayer nodes}, in which multiple players have legal decisions available; and \emph{terminal state nodes}, which correspond to terminal states and have outcomes assigned to them.

A game tree generated by \cref{alg:GameTree} is a sort of hybrid between game theoretic extensive- and strategic-form games (e.g., see \cite{rasmusen_games_2007}), except with no information sets.
On single-player nodes, individual players choose an outgoing edge to follow. 
To capture simultaneous play, multiplayer nodes act as strategic (``matrix'') form games: several players must simultaneously make a decision, which is then evaluated by an umpire to choose an outgoing edge to follow.
(We describe this strategic-form game with a \emph{decision matrix}, see \cref{def:DecisionMatrix}.)
This avoids the ambiguities inherent in the information set construction of simultaneous play in extensive form games \cite{bonanno_set-theoretic_1992}, which relies on sequential moves with hidden information, and can be experimentally different from true simultaneous play \cite{cooper_communication_1992,salles_beyond_2008}.
Ultimately, an outcome is produced at a terminal node.

\section{Agency Equivalence}
\label{sec:AgencyEquivalence}

The game system \cref{def:UnderlyingGameSystem} is deliberately flexible, in order to describe any discrete game.
However, its flexibility means that there are several ways to express the ``same'' game.
Here we develop two precise senses of this ``sameness'': \emph{game tree equivalence (up to relabeling)} and \emph{agency equivalence}.
This is a useful precursor to measuring distance between games, where we might wish equivalent games to have zero distance.

Game tree equivalence up to relabeling (\cref{def:GameTreeEquivalenceUpToRelabeling} and \cref{sec:GameTreeEquivalenceUpToRelabeling}) matches game systems if they produce the same game trees, with some variation in aesthetic labeling: e.g., the names of decisions or states can differ, but probabilities cannot. 

Agency equivalence (\cref{def:AgencyEquivalence}) matches game systems if they offer players the same agency, that is, the same sorts of meaningful choices with the same sorts of consequences.
We will define this by performing a series of reductions on game trees to prune spurious differences, declaring two game systems equivalent if their reduced trees match.
There are four kinds of differences that we consider spurious for this purpose; we describe them heuristically here, then formally in \cref{sec:GameTreeReductions}:

A \emph{bookkeeping subtree} (\cref{def:BookkeepingSubtree}) is a portion of a game tree where there is only one decision available at each state.
Though there may be randomness involved, play continues on the subtree inevitably, without any chance for player influence.
Consider a version of tic-tac-toe, in which players must declare end-of-turn after placing a symbol (X, end-of-turn, O, end-of-turn, ...).
From the standpoint of player agency, we do not consider this meaningfully different from standard tic-tac-toe, in which turns automatically advance (X, O, X, O, ...).

A \emph{single-player subtree} (\cref{def:SinglePlayerSubtree}) is a portion of the game tree where the same single player makes several deterministic decisions in a row.
There is no difference in options or outcomes if the player makes these decisions one at a time or all at once.
For instance, pawn promotion in chess could be split into two steps with an intermediate state (move, then promote), or lumped into one (move-and-promote). 
We do not consider these different from the standpoint of player agency.

A \emph{symmetry-redundant subtree} (\cref{def:SymmetryRedundantSubtree}) is a portion of the game tree that is unnecessary because it duplicates a sibling subtree.
Because of the board symmetry in tic-tac-toe, starting in one corner versus another corner leads to substantively the same remaining decisions for the rest of the game---even though the precise game states are different.
Even if a version of tic-tac-toe forbid players from starting in three of the corners, and three of the sides, we would consider it identical to standard tic-tac-toe from the standpoint of meaningful player agency.

Finally, a \emph{decision matrix redundancy} (\cref{def:DecisionMatrixRedundancy}) occurs when a player has two decisions at a state that would have identical results---it really does not matter which one they pick.
This player would have the same agency if they had only one of those decisions available.
Putting all these together:
\begin{definition}
    \label{def:AgencyEquivalence}
    We say two game systems $\mathcal{G}$ and $\mathcal{G}'$ are \emph{agency equivalent} if their respective game trees%
    \footnote{For now, this and following definitions can only be usefully applied to finite game trees, though infinite trees could be truncated and similarly compared.}
    can be made equivalent up to relabeling (\cref{def:GameTreeEquivalenceUpToRelabeling}) by performing the following reductions, as many times as necessary, in any order:
    \begin{itemize}
        \item Bookkeeping subtree reduction (\cref{def:BookkeepingSubtree})
        \item Single-player subtree reduction (\cref{def:SinglePlayerSubtree})
        \item Symmetry-redundant subtree reduction (\cref{def:SymmetryRedundantSubtree})
        \item Decision matrix redundancy reduction (\cref{def:DecisionMatrixRedundancy})
    \end{itemize}
\end{definition}

The following subsections flesh out the technical details for these two senses of equivalence.
Note, the reductions mentioned in \cref{def:AgencyEquivalence} produce \emph{reduced game trees}, editing the game trees produced from \cref{alg:GameTree} to remove non-essential information.
The term ``game tree'' should be understood below to refer to both reduced and unreduced game trees.

\subsection{Game Tree Equivalence up to Relabeling}
\label{sec:GameTreeEquivalenceUpToRelabeling}

To start, let's delete all tree labels and match what remains:

\begin{definition}
    \label{def:StrippedGameTree}
    A \emph{stripped game tree}, denoted $\langle T \rangle$, is a game tree $T$ with all labels removed; only the arrangement of nodes and edges remains.
\end{definition}

\begin{definition}[Structural equivalence]
    \label{def:StructuralEquivalence}
    Two game trees $T,T'$ (or game systems $\mathcal{G},\mathcal{G}'$) are \emph{structurally equivalent} if the stripped trees are equal $\langle T \rangle = \langle T' \rangle$ (or if the sets of stripped game trees are equal $\langle \tau(\mathcal{G}) \rangle = \langle \tau(\mathcal{G}') \rangle$). 

    This establishes a bijective \emph{structural correspondence} $f: n \mapsto n'$, similarly $f: e \mapsto e'$, between the labelled nodes and edges of corresponding trees $T$ and $T'$ (or $t \in \tau(\mathcal{G})$ and $t' \in \tau(\mathcal{G}')$).
    Several such correspondences may be possible (e.g., because a tree is symmetric).
\end{definition}

This is sufficient to say that the arrangement of nodes and edges is the same.
However, some labels do contain important content that distinguishes two game systems in substance, not just aesthetics.
In particular, we want to see that corresponding probabilities are the same, players have the same kinds of choices available, and that the outcomes are similarly distinct.  
Comparing probabilities and outcomes is straightforward:
\begin{definition}[Matching probabilities]
    \label{def:CorrespondingProbabilities}
    For each chance edge $e$ in a game tree, let $\p(e)$ be the assigned probability.
    Two structurally equivalent game trees $T,T'$ (or game systems $\mathcal{G}, \mathcal{G}'$) with structural correspondence $f: T \to T'$ are said to have \emph{matching probabilities} if $\p(e) = \p(f(e))$ for all chance edges $e \in T$ ($\in \tau(\mathcal{G})$).
\end{definition}

\begin{definition}[Similarly distinct outcomes]
    \label{def:CorrespondinglyDistinctOutcomes}
    Take two structurally equivalent game trees $T,T'$ (or game systems $\mathcal{G},\mathcal{G}'$) with correspondence $f$, let $O, O'$ be the sets of all distinct outcomes assigned to their respective terminal nodes, and let $\Omega(z)$ be the outcome assigned to a terminal node $z$.
    We say $T$ and $T'$ (or $\mathcal{G}$ and $\mathcal{G}'$) have \emph{similarly distinct outcomes} if there exists a bijective map $o: O \to O'$ such that $\Omega(z) = o(\Omega(f(z)))$ for all terminal nodes $z \in T$ ($\in \tau(\mathcal{G})$).
\end{definition}

Confirming that players have the same kinds of decisions along the way is more involved, at least formally.
We want single-player nodes to still be single-player nodes with the same number of choices, and multiplayer nodes to still be multiplayer nodes with the same interaction between each player's simultaneous decisions.
In essence, we want the same strategic-form game to be played at each internal state node, as described in \cref{sec:GameTreesAndAutomata}.
First let us define the decision matrix, which describes these strategic (``matrix'') form games:
\begin{definition}[Decision matrix]
    \label{def:DecisionMatrix}
    Let $\mathcal{G}$ be a game system with players $\mathcal{P} = ( p_1, \ldots, p_n )$.
    Let $w$ be a non-terminal state node in a game tree $T \in \tau(\mathcal{G})$ with assigned state $s$ and outgoing decision edges $E$.
    Each player $p$ has a set of legal choices $\ell_p(w)$ they can select to influence the edge followed.
    Let $\ell^0_p(w) = \ell_p(w)$ unless $\ell_p(w)$ is empty, in which case $\ell^0_p(w) = \{0\}$, with 0 the null choice.
    The \emph{decision matrix} at node $w$ is a map $D_w: \ell^0_{p_1}(w)\times\cdots\times \ell^0_{p_n}(w)\to E$ of decision tuples to edges.

    A game tree produced freshly from \cref{alg:GameTree} has $\ell_p(w) = L_p(s)$, the usual legal set (see \cref{def:UnderlyingGameSystem}), but game tree reductions or transformations may adjust $\ell$ (e.g., see \cref{def:SinglePlayerSubtree,def:DecisionMatrixRedundancy}).
\end{definition}

In a tree with $\ell_p(w) = L_p(s)$ and each edge $e \in E$ labeled with a set of one or more unique decision tuples, $D_w$ simply maps each decision tuple $(d_{p_1}, \ldots, d_{p_n}) \in \mathcal{D}_0^n$ to the edge with that tuple.
An edge might obtain multiple tuples, even though \cref{alg:GameTree} only assigns one to each edge, due to something like a symmetry-redundant subtree reduction (see \cref{def:SymmetryRedundantSubtree}).

For example, \cref{fig:DecisionMatrixIllustration} is a sample decision matrix with outgoing edges as it might appear in a reduced game tree, for a 3-player game with $\mathcal{P} = (\text{P1},\text{P2},\text{P3})$. 
The alternative labeling \cref{fig:DecisionMatrixIllustrationRelabeled} helps clarify the structure of the joint decisions.
Omitting the inactive P2, who has no legal choices:
\vspace{-15pt}
\begin{figure}[H]
\captionsetup{skip=0pt}
\centering
\subfloat{
    \hspace{-5pt}
    \includegraphics[scale=1,trim=1 12 13 3,clip]{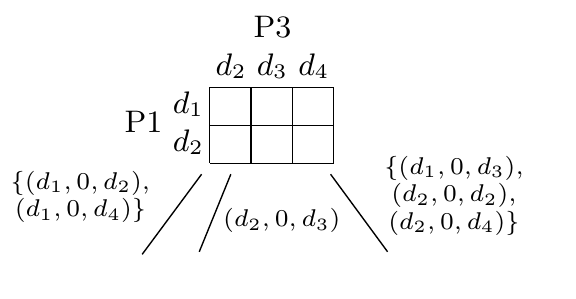}
    \hspace{3pt}
\label{fig:DecisionMatrixIllustration}}
\subfloat{
    \includegraphics[scale=1,trim=10 12 26 3,clip]{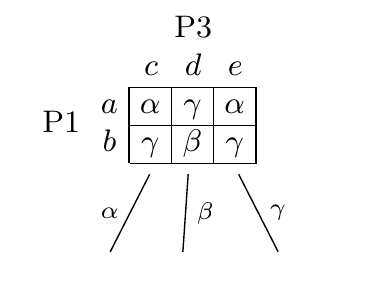}
\label{fig:DecisionMatrixIllustrationRelabeled}}
\caption{Left: An example decision matrix in a reduced game tree
(e.g., after reduction by \cref{def:SinglePlayerSubtree} or \cref{def:SymmetryRedundantSubtree}).
Right: The same matrix, relabeled.} 
\label{fig:DecisionMatrixIllustrationAndRelabeled}
    \vspace{-5pt}
\end{figure}

We say that two decision matrices match (in the context of a game tree) if there is a self-consistent way to relabel the players, decisions, and edges such that the relabeled players making the relabeled decisions lead to the relabeled edges:
\begin{definition}[Matching decision matrices]
    \label{def:MatchingDecisionMatrices}
    Take two game trees $T$ and $T'$ that are structurally equivalent with correspondence $f$.
    Let $w \in T$ and $w' = f(w) \in T'$ be corresponding non-terminal state nodes, with states $s, s'$, decision matrices $D_w,D_{w'}$, and sets of players $\mathcal{P}, \mathcal{P}'$, respectively.

    If possible, define a bijective \emph{player correspondence} $\pi: \mathcal{P} \to \mathcal{P}'$, with associated bijective maps $\lambda_{\pi,p}: \ell^0_p(w) \to \ell^0_{\pi(p)}(w')$ ($p \in \mathcal{P}$), which correspond choices in the two games.
    Together, these furnish a unique map between the decision tuples, $\lambda_\pi: d_0^n \mapsto (d_0^n)'$, i.e., between the domains of $D_w$ and $D_{w'}$.

    The decision matrices $D_w$ and $D_{w'}$ are said to \emph{match}, denoted $D_w \sim D_{w'}$, if there exists at least one such $\pi$ and set $\{\lambda_{\pi,p}\}$ such that the matrices map to corresponding edges: i.e., $D_w(d_0^n) = e \in T$ and $D_{w'}(\lambda_\pi(d_0^n)) = f(e) \in T'$ for all $d_0^n$ in the domain of $D_w$.
\end{definition}

See \cref{fig:MatchingDecisionTreesExample} for examples.
\cref{fig:DecisionMatrixIllustration,fig:DecisionMatrixIllustrationRelabeled} could also be said to match, since they only differ in choice labeling, if the node and edges were placed to correspond in two structurally equivalent trees.
Note that we may relabel the decisions for each player separately (e.g., $d_2 \to b$ for P1 in \cref{fig:DecisionMatrixIllustrationAndRelabeled}, but $d_2 \to c$ for P3), and we may even relabel the null decision.

\setlength{\abovecaptionskip}{2pt}
\newcommand\bigredsim{\stackrel{\mathclap{\normalfont\mbox{\footnotesize agency}}}{\scalebox{3.5}{$\sim$}}}
\newcommand\redsim{\mathrel{\overset{\makebox[0pt]{\mbox{\normalfont\tiny red.}}}{\sim}}}
\begin{figure*}[!tb]
    \centering
    \vspace{-3pt}
    \setlength\tabcolsep{8pt}
    \begin{tabular}{m{.36\linewidth} m{.04\linewidth} m{0.40\linewidth}}
        \includegraphics[scale=1,trim=3 0 15 0,clip]{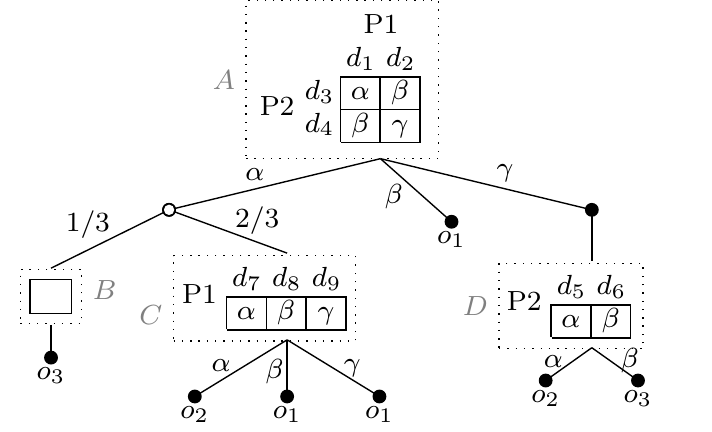}
    &
    $\bigredsim$
    &
        \hspace{3pt}
    \includegraphics[scale=1,trim=3 0 15 0,clip]{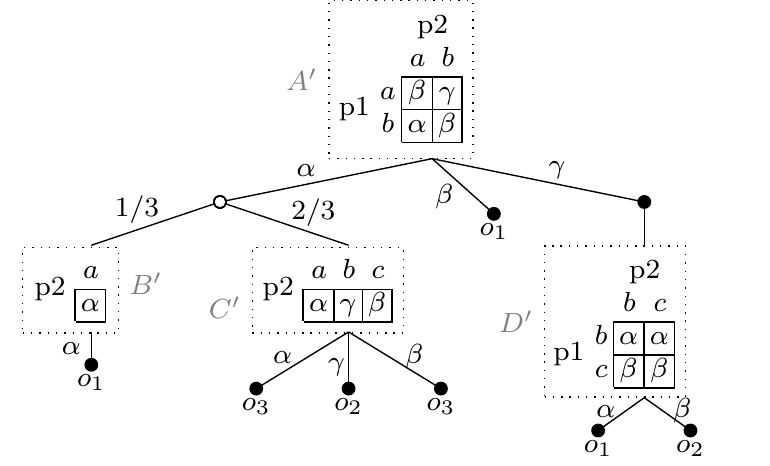}
    \end{tabular}
    \vspace{-2pt}
    \caption{
        These are two full game trees, with state labels suppressed, outcomes $o_i$ on each terminal node, probabilities ($1/3$ and $2/3$) on chance edges, and decision edges labeled with decision matrix outcomes instead of sets of decision tuples, for brevity (e.g., as \cref{fig:DecisionMatrixIllustrationRelabeled} relabels \cref{fig:DecisionMatrixIllustration}).
        Decision matrices are illustrated on all non-terminal state nodes, with inactive players not shown (like in \cref{fig:DecisionMatrixIllustration}).
        Some (but not all) of the decision matrices match: $A \sim A'$, $B \not\sim B'$, $C \sim C'$, and $D \not\sim D'$.
        However, all of them match after reduction by \cref{def:DecisionMatrixRedundancy}: in particular, $B \redsim B'$ and $D \redsim D'$ (the blank matrix $B$ is a decision matrix with empty domain, see \cref{def:DecisionMatrixRedundancy}).
        In fact, $B$ ($D$) is the reduced form of $B'$ ($D'$), after some relabeling.
        Thus, after these reductions, the \emph{trees have matching decision matrices} (\cref{def:TreesWithMatchingDecisionMatrices}) under the player correspondence P1 $\leftrightarrow$ p2 and P2 $\leftrightarrow$ p1.
        In fact, because the probabilities also match and the outcomes are similarly distinct, the two trees are agency equivalent (\cref{def:AgencyEquivalence}).
    }
    \label{fig:MatchingDecisionTreesExample}
    \vspace{-5pt}
\end{figure*}

We can now generalize beyond individual decision matrices to game trees and systems
(see \cref{fig:MatchingDecisionTreesExample}).
Whatever relabeling is necessary to make the decision matrices match, we demand at least that the player relabeling is the same everywhere.
It is unimportant if the decision labels vary from matrix to matrix.
\begin{definition} \textbf{(Trees with matching decision matrices)}
    \label{def:TreesWithMatchingDecisionMatrices}
    Take two structurally equivalent game trees $T,T'$ (or game systems $\mathcal{G},\mathcal{G}'$) with structural correspondence $f: T \to T'$ and sets of players $\mathcal{P}$ and $\mathcal{P}'$. 
    We say these trees (or systems) have \emph{matching decision matrices} if there exists at least one bijective player correspondence $\pi: \mathcal{P} \to \mathcal{P}'$ such that all corresponding decision matrices match with respect to $\pi$---%
    i.e., $D_w \sim D_{f(w)}$ for all internal state nodes $w \in T\ ( \in \mathcal{G})$ with $\pi$ as the player correspondence. (See \cref{def:MatchingDecisionMatrices}.
    The associated decision mappings $\{\lambda_{\pi,p}\}$ may be different for each node in the tree.)
\end{definition}

Finally, let us put all of this together to give a broadly useful sense of equivalence between game trees, which respects everything about them except for the specific labels chosen to represent players, states, decisions, and outcomes:
\begin{definition}[Equivalence up to relabeling]
    \label{def:GameTreeEquivalenceUpToRelabeling}
    We say that two game trees $T,T'$ are \emph{equivalent up to relabeling} (or that game systems $\mathcal{G},\mathcal{G}'$ are \emph{game tree equivalent up to relabeling}) if $T$ and $T'$ (or $\mathcal{G}$ and $\mathcal{G}'$) are structurally equivalent and have matching probabilities, matching decision matrices, and similarly distinct outcomes, all with respect to the same structural correspondence $f$.
\end{definition}

If all labels additionally happen to be identical, then the game trees (or game systems) are simply \emph{equivalent} (or \emph{game tree equivalent}).
If only some labels are additionally identical, we might say (using outcomes as an example) that two trees $t$ and $t'$ are equivalent up to relabeling and with identical outcomes.
This means that if $t$ and $t'$ have structural correspondence $f: t \to t'$ and $\Omega(z)$ gives the outcome assignment of terminal node $z \in t$, then $\Omega(z) = \Omega(f(z))$ for all $z$.

It is worth noting that \cref{def:GameTreeEquivalenceUpToRelabeling}, in not distinguishing between the content of outcome labels, does not distinguish whether an outcome might be good or bad for a player.
The normal and \emph{mis\`ere} versions of a game have opposite win/lose conditions, for instance, but would be considered equivalent up to relabeling.

\subsection{Game Tree Reductions}
\label{sec:GameTreeReductions}

To establish agency equivalence from \cref{def:AgencyEquivalence}, we need to prune those differences between trees that are not meaningful from the standpoint of player agency.
Here we describe the relevant transformations to reduce bookkeeping subtrees, single-player subtrees, symmetry-redundant subtrees, and decision matrix redundancies, as heuristically described in \cref{sec:AgencyEquivalence}.

\begin{figure*}[!tb]
    \vspace{-2pt}
    \centering
    \begin{tabular}{m{.42\linewidth} m{.06\linewidth} m{0.42\linewidth}}
        \includegraphics[scale=0.95,trim=0 7 0 7,clip]{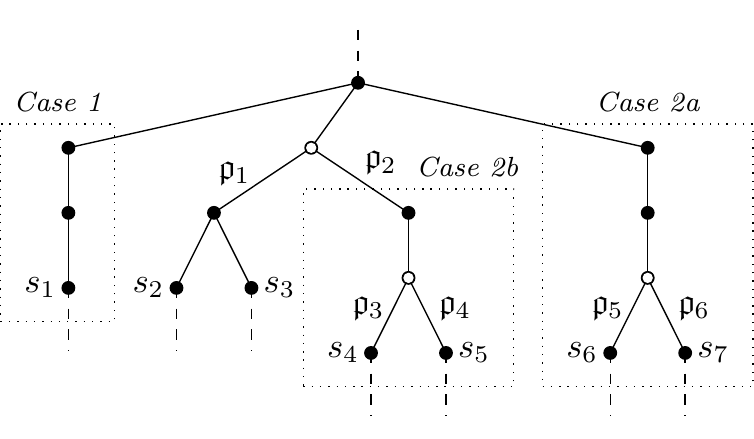}
    &
    \tikz \draw[-{latex},line width=1mm] (0,0) -- (1,0);
    &
    \includegraphics[scale=0.95,trim=0 7 0 7,clip]{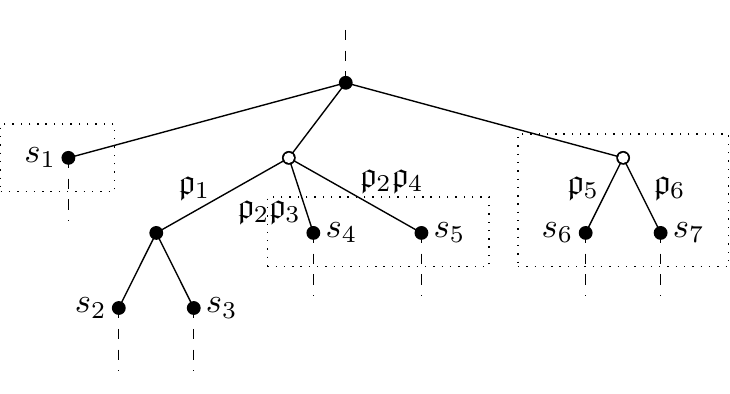}
    \end{tabular}
    \caption{
        Bookkeeping subtree reduction example. 
        Dotted lines highlight the bookkeeping subtrees before and after being reduced, exemplifying the different cases in \cref{def:BookkeepingSubtree}.
        Solid dots are state nodes, circles are chance nodes, and chance edges are labeled with probabilities $\p_i$.
        The labels on most state nodes and all decision edges have been omitted.
        Dashed lines connect to other parts of the game tree.
    }
    \label{fig:BookkeepingReductionExample}
    \vspace{-5pt}
\end{figure*}

\begin{definition}
    \label{def:BookkeepingSubtree}
    A \emph{bookkeeping subtree} is a subtree of a game tree rooted at a state node $r$ and with state nodes as leaves, which has exactly one decision edge proceeding out of $r$ and each interior state node.
    Players cannot influence play in this subtree.
    Such a subtree may be reduced as follows (see \cref{fig:BookkeepingReductionExample}):\medskip

    \emph{Case 1:} If there are no chance nodes in the subtree:
        \begin{enumerate}
            \item There is only a single leaf, with state $s$.
                Replace the entire subtree with a state node with state $s$.
        \end{enumerate}
        \smallskip

    \emph{Case 2:} If there are chance nodes in the subtree:
    \begin{enumerate}
        \item Let $G$ be the set of all paths from $r$ to the subtree leaves.
            Let $l(g)$ be the final state node in each $g\in G$ (i.e., the leaves).
        \item Assign each path $g$ a probability $\p(g)$ given by the product of the probabilities on the chance edges in $g$.
        \item \label{step:DefineNodeC} Then, if the parent $r'$ of the subtree root $r$ is \mydots
            \begin{enumerate}[leftmargin=4mm]
                \item \emph{Case 2a:} \mydots a state node: 
                    Replace $r$ with a new chance node $c$.
                \item \emph{Case 2b:} \mydots a chance node $c$\,: proceed to step \ref{step:DeleteAllNodes}. 
                    (Note $r$ has an incoming chance edge with probability $\p_r$.)
                \item \emph{Case 2c:} \mydots nonexistent ($r$ is the root of the game tree): 
                    Replace the child of $r$ with a new chance node $c$.
            \end{enumerate}
        \item \label{step:DeleteAllNodes} Taking the chance node $c$ from step \ref{step:DefineNodeC}, delete all nodes and edges between $c$ and the leaves, non-inclusive.
        \item Draw new chance edges between $c$ and each leaf $l(g)$, labeled by the corresponding probabilities $\p(g)$, or $\p_r\cdot \p(g)$ in Case 2b.
    \end{enumerate}
\end{definition}

\begin{figure*}[tb!]
    \vspace{-2pt}
    \centering
    \begin{tabular}{m{.38\linewidth} m{.03\linewidth} m{0.45\linewidth}}
    \includegraphics[scale=0.90,trim=1 7 1 7,clip]{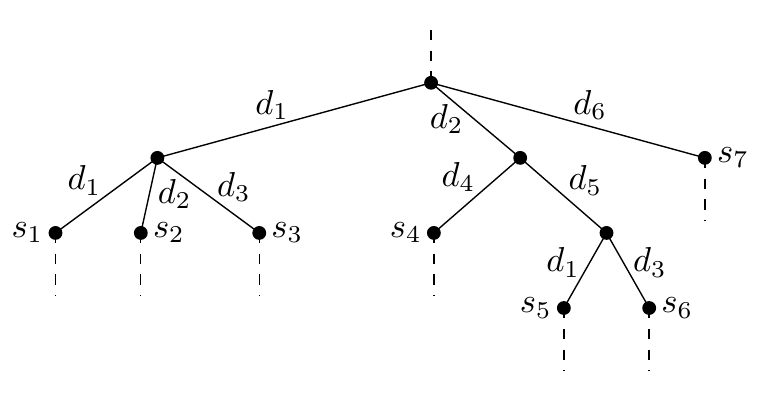}
    &
    \tikz \draw[-{latex},line width=1mm] (0,0) -- (1,0);
    &
    \includegraphics[scale=0.90,trim=1 7 1 7,clip]{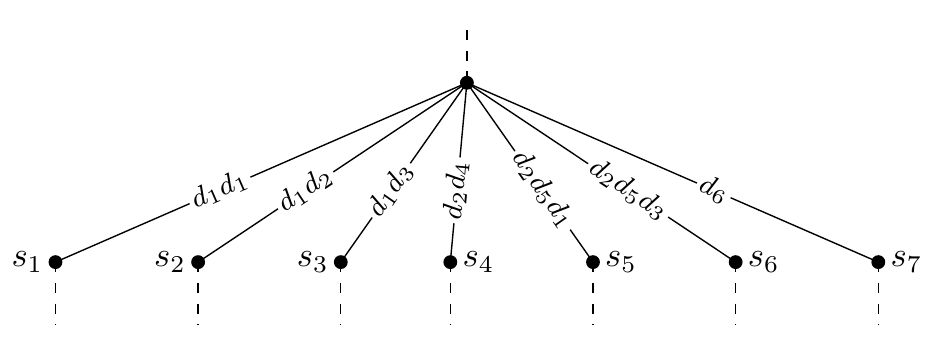}
    \end{tabular}
    \caption{
        Single-player subtree reduction example, illustrating \cref{def:SinglePlayerSubtree}.
        All state nodes, except possibly the leaves, belong to the same player.
        Internal state node labels have been omitted.
        Edges have been labeled only with this player's decisions, for brevity, since all other players take the null decision everywhere.
        Dashed lines connect to other parts of the game tree.
    }
    \label{fig:SinglePlayerReductionExample}
    \vspace{-7pt}
\end{figure*}

\begin{definition}
    \label{def:SinglePlayerSubtree}
    A \emph{single-player deterministic subtree} is a subtree of a game tree rooted at a state node $r$ and with state nodes as leaves, without any chance nodes, in which all nodes belong to a single player (except perhaps the leaves).
    Only that player has any meaningful decisions in this subtree, and they could just as well be made all at once.
    Such a subtree may be reduced as follows (see \cref{fig:SinglePlayerReductionExample}):
    \begin{enumerate}
        \item Let $G$ be the set of all paths from $r$ to its leaves.
            Let $l(g)$ be the final state node in each $g\in G$ (i.e., the leaves).
        \item Delete all nodes and edges between $r$ and the leaves, non-inclusive.
        \item Draw a new decision edge between $r$ and each leaf $l(g)$, labeled by the sequence of decision tuples in $g$.
    \end{enumerate}

    This reduction also changes the domain of the decision matrix $D_r$ at $r$ (see \cref{def:DecisionMatrix}):
    The legal choices $\ell_p(r)$ available to the single player $p$ at $r$ are now the set of decision tuple sequences from each $g \in G$ (i.e., the labels on the new decision edges proceeding from $r$), \emph{not} the canonical legal set $L_p$.
\end{definition}

\begin{definition}
    \label{def:SymmetryRedundantSubtree}
    A \emph{symmetry-redundant subtree} is a subtree $t$ of a game tree rooted at a state or chance node $r$ and extending to all its descendants, that is equivalent up to relabeling, and with identical players and outcomes, to a subtree $t'$ rooted at a sibling node $r'$ (and extending to all of its descendants).
    (A lone terminal node $z$ is also considered a symmetry-redundant subtree if its outcome $\Omega(z)$ is identical to one of its siblings.)
    The symmetry-redundant subtree $t$ may be reduced as follows:
    \begin{enumerate}[leftmargin=6mm]
        \item Let $e,e'$ be the edges ingoing to $r,r'$.
            If $e$ and $e'$ are \ldots
            \begin{enumerate}[leftmargin=3mm]
                \item \emph{Case 1:} \ldots decision edges with assigned tuple sets $d(e)$ and $d(e')$ (if only one tuple, consider it a set of size 1).
                    Replace the tuple set on $e'$ with $d(e')\cup d(e)$.
                \item \emph{Case 2:} \ldots chance edges with assigned probabilities $\p(e)$ and $\p(e')$:
                    Replace the probability on $e'$ with $\p(e') + \p(e)$.
            \end{enumerate}
        \item Delete the entire subtree $t$ rooted at $r$, and the edge $e$.
        \item (For \emph{Case 2:}) If $e'$ now has assigned probability 1, this chance edge is superfluous, as is the chance node parent $c'$ of $r'$. 
            Delete $e'$, and replace $c'$ by moving $r'$ (with subtree attached) into its place.
    \end{enumerate}
\end{definition}
Note symmetry-redundant subtrees may commonly occur when the sibling nodes $r$ and $r'$ are assigned the same state, e.g., when two different decisions lead to the exact same game state.

Several decision tuples may end up on a single edge (e.g., \cref{fig:DecisionMatrixIllustration}) because of symmetry-redundant subtree reductions, which may lead to redundancies in decision matrices---if they were not redundant already.
Consider \cref{fig:DecisionMatrixIllustrationRelabeled}, for instance, which relabels \cref{fig:DecisionMatrixIllustration}. 
It is clear that P3 gains no additional agency by having choice $e$ available in addition to choice $c$.
We can eliminate such meaningless redundancies in decision matrices by removing duplicate rows and columns:
\begin{definition}
    \label{def:DecisionMatrixRedundancy}    
    A \emph{decision matrix redundancy} occurs when a decision matrix $D_w: \ell^0_{p_1}\times\cdots\times\ell^0_{p_n} \to E$ contains more than one choice for some player $p$ which lead to the same result.
    That is, when there exist distinct choices $a,b \in \ell_{p_i}$ for some $p_i$ such that  $D_w(d_1,\ldots,d_{i-1},a,d_{i+1},\ldots,d_k) = D_w(d_1,\ldots,d_{i-1},b,d_{i+1},\ldots,d_k)$ for all possible choices $d_j \in \ell_{p_j}$.
    The choices $a$ and $b$ are redundant.

    To eliminate redundancies and unnecessary bookkeeping distinctions, a \emph{reduced decision matrix} can be produced as follows (see \cref{fig:MatchingDecisionTreesExample} for examples):
    \begin{enumerate}
        \item If there exist two redundant choices $a,b \in \ell_{p_i}$ for some $p_i$, delete one: $\ell_{p_i} \to \ell_{p_i} \setminus \{ b \}$.
        Repeat until no redundancies remain for any player.
    \item If all players have only a single (possibly null) choice remaining ($|\ell^0_{p}| = 1$), there must only be a single edge $e$ in the image of $D_w$.
        We may define $D_w: \varnothing \to \{ e \}$.
    \end{enumerate}
    Any two corresponding decision matrices with empty domains are said to match in the sense of \cref{def:MatchingDecisionMatrices}.
Step 2 is not strictly necessary for establishing equivalence, but reflects that in a bookkeeping subtree, it does not matter which player(s) are given the task of executing the bookkeeping.

\end{definition}

\section{Discussion: Towards Game Similarity}
\label{sec:Discussion}

We have proposed a grammar-like formalism to describe finite discrete game systems without hidden information, along with equivalence relations on this space of games, that are insensitive to cosmetic variations in game rules.
Developing measures of game equivalence and similarity will be important for formally interrogating the design of games, and as a steppingstone towards predicting player behavior from design patterns.
More broadly, we hope that such efforts may help connect game design and mathematical experts, enriching the many applications of games by exploring formal analogues to the rich tools and vocabulary used in game design today.

However, for complex games it may be impractical to check equivalence by drawing, reducing, and comparing full game trees.
One possible way forward is to learn to transform and compare the grammars directly, using \cref{def:AgencyEquivalence,def:GameTreeEquivalenceUpToRelabeling} as guidance for what those transformations must accomplish.

We could also move beyond game equivalence, to consider game similarity.
Suppose a ludologer supplies a mapping $\psi$ between the state spaces $\mathbb{S}, \mathbb{S}'$ (and perhaps also players and outcomes) of two game systems $\mathcal{G}, \mathcal{G}'$.
Then the systems could be compared by sampling states $s \in \mathbb{S}$, computing some function (e.g., a partial game tree) at $s$ and $\psi(s) \in \mathbb{S}'$, comparing the function values (e.g., assigning 1 if the partial trees are equivalent after appropriate reductions, 0 otherwise), and averaging the results.
This would give a quantitative measure of game similarity (e.g., between 0 and 1), and another way to check equivalence (e.g., if similarity $= 1$).
Even if only a fraction of the states are randomly sampled, a confidence interval could be estimated for the computed similarity.

There are several details to work out here in the calculation, interpretation, and likelihood estimates of such a similarity measure, which we leave for future work.
For instance: in contrast to comparing full game trees, this similarity method could sample many states not legally accessible in standard play, so it would compare game rules beyond just legal gameplay.
Also, since state spaces for complex games can be gargantuan, it is unclear how quickly a similarity estimate would converge.
Nevertheless, when combined with the intuitive and technical guidance of the equivalence relations \cref{def:AgencyEquivalence} and \cref{def:GameTreeEquivalenceUpToRelabeling}, such a sampling method has the potential to practically measure distances between games, without sensitivity to cosmetic variations in the rule descriptions, and with minimal input from ludologers.
We look forward to exploring this and other game similarity measures in future work.

\section*{Acknowledgments}

We thank
Stephen Crane,
Marquita Ellis,
Ryan Janish, 
Will Johnson, 
Kiran Lakkaraju,
Kweku Opoku-Agyemang, 
Stephen Phillips, 
and
Ben Wormleighton
for useful discussions.
We also thank Vlaada Chv\'atil for designing \emph{Mage Knight} \cite{chvatil_mage_2011}, the board game that inspired this research.
This work is supported in part by the Office of Naval Research under the Embedded Humans MURI (N00014-13-1-0341) as well as a Philippine- California Advanced Research Institutes (PCARI) grant.

\bibliographystyle{IEEEtran.bst}
\bibliography{priggins-FGGE-2020}

\begin{thebibliography}{10}
\providecommand{\url}[1]{#1}
\csname url@samestyle\endcsname
\providecommand{\newblock}{\relax}
\providecommand{\bibinfo}[2]{#2}
\providecommand{\BIBentrySTDinterwordspacing}{\spaceskip=0pt\relax}
\providecommand{\BIBentryALTinterwordstretchfactor}{4}
\providecommand{\BIBentryALTinterwordspacing}{\spaceskip=\fontdimen2\font plus
\BIBentryALTinterwordstretchfactor\fontdimen3\font minus
  \fontdimen4\font\relax}
\providecommand{\BIBforeignlanguage}[2]{{%
\expandafter\ifx\csname l@#1\endcsname\relax
\typeout{** WARNING: IEEEtran.bst: No hyphenation pattern has been}%
\typeout{** loaded for the language `#1'. Using the pattern for}%
\typeout{** the default language instead.}%
\else
\language=\csname l@#1\endcsname
\fi
#2}}
\providecommand{\BIBdecl}{\relax}
\BIBdecl

\bibitem{yannakakis_artificial_2018}
G.~N. Yannakakis and J.~Togelius, \emph{Artificial Intelligence and
  Games}.\hskip 1em plus 0.5em minus 0.4em\relax Springer, 2018.

\bibitem{melcer_games_2015}
E.~Melcer, T.-H.~D. Nguyen, Z.~Chen, A.~Canossa, M.~S. El-Nasr, and
  K.~Isbister, ``Games research today: Analyzing the academic landscape
  2000-2014,'' in \emph{Proc. 10th Int. Conf. Foundations of Digital Games},
  2015.

\bibitem{melcer_toward_2017}
E.~Melcer and K.~Isbister, ``Toward understanding disciplinary divides within
  games research,'' in \emph{Proc. Int. Conf. Foundations of Digital
  Games}.\hskip 1em plus 0.5em minus 0.4em\relax {ACM} Press, 2017.

\bibitem{tekinbas_rules_2003}
K.~S. Tekinbaş and E.~Zimmerman, \emph{Rules of play: game design
  fundamentals}.\hskip 1em plus 0.5em minus 0.4em\relax {MIT} Press, 2003.

\bibitem{koster_theory_2005}
R.~Koster, \emph{A theory of fun for game design}.\hskip 1em plus 0.5em minus
  0.4em\relax Paraglyph Press, 2005.

\bibitem{adams_game_2012}
E.~Adams and J.~Dormans, \emph{Game mechanics: advanced game design}.\hskip 1em
  plus 0.5em minus 0.4em\relax New Riders, 2012.

\bibitem{schell_art_2019}
J.~Schell, \emph{The art of game design: a book of lenses}, 3rd~ed.\hskip 1em
  plus 0.5em minus 0.4em\relax Taylor \& Francis, 2019.

\bibitem{engelstein_building_2019}
G.~Engelstein and I.~Shalev, \emph{Building blocks of tabletop game design: an
  encyclopedia of mechanisms}.\hskip 1em plus 0.5em minus 0.4em\relax Taylor \&
  Francis, 2019.

\bibitem{riggins_tools_2019}
P.~Riggins and D.~{McPherson}, ``Tools for mathematical ludology,'' 2019,
  {arXiv}:1912.03295 [cs.AI].

\bibitem{cousins_elementary_2004}
B.~Cousins, ``Elementary game design,'' \emph{Develop Magazine}, pp. 51--54,
  2004.

\bibitem{koster_grammar_2005}
R.~Koster, ``A grammar of gameplay,'' 2005, {G}ame {D}evelopers {C}onf.

\bibitem{stephane_game_2006}
\BIBentryALTinterwordspacing
B.~Stéphane. (2006) A game grammar. [Online]. Available:
  \url{http://www.stephanebura.com/diagrams/}
\BIBentrySTDinterwordspacing

\bibitem{browne_modern_2018}
C.~Browne, ``Modern techniques for ancient games,'' in \emph{2018 {IEEE} Conf.
  Computational Intelligence and Games ({CIG})}.\hskip 1em plus 0.5em minus
  0.4em\relax {IEEE}, 2018.

\bibitem{cooper_communication_1992}
R.~Cooper, D.~V. {DeJong}, R.~Forsythe, and T.~W. Ross, ``Communication in
  coordination games,'' \emph{Q. J. Econ.}, vol. 107, no.~2, pp. 739--771,
  1992.

\bibitem{salles_beyond_2008}
P.~J. Hammond, ``Beyond normal form invariance: First mover advantage in
  two-stage games with or without predictable cheap talk,'' in \emph{Rational
  Choice and Social Welfare}.\hskip 1em plus 0.5em minus 0.4em\relax Springer,
  2008, pp. 215--233.

\bibitem{beck_combinatorial_2008}
J.~Beck, \emph{Combinatorial games: tic-tac-toe theory}.\hskip 1em plus 0.5em
  minus 0.4em\relax Cambridge University Press, 2008.

\bibitem{rasmusen_games_2007}
E.~Rasmusen, \emph{Games and information: an introduction to game theory},
  4th~ed.\hskip 1em plus 0.5em minus 0.4em\relax Blackwell Pub, 2007.

\bibitem{piette_ludii_2019}
{\'E}.~Piette, D.~J. N.~J. Soemers, M.~Stephenson, C.~F. Sironi, M.~H.~M.
  Winands, and C.~Browne, ``Ludii - the ludemic general game system,'' 2019,
  arXiv:1905.05013 [cs.AI].

\bibitem{thielscher_general_2010}
M.~Thielscher, ``A general game description language for incomplete information
  games,'' in \emph{Proc. 24th {AAAI} Conf. Artificial Intelligence}, ser.
  {AAAI}'10.\hskip 1em plus 0.5em minus 0.4em\relax {AAAI} Press, 2010, pp.
  994--999.

\bibitem{love_general_2006}
N.~Love, T.~Hinrichs, and M.~Genesereth, ``General game playing: Game
  description language specification,'' Stanford Logic Group, Computer Science
  Department, Tech. Rep. {LG}-2006-01, 2006.

\bibitem{thielscher_gdl-iii:_2017}
M.~Thielscher, ``{GDL}-{III}: A description language for epistemic general game
  playing,'' in \emph{Proc. 26th Int. Joint Conf. Artificial Intelligence},
  ser. {IJCAI}'17.\hskip 1em plus 0.5em minus 0.4em\relax {AAAI} Press, 2017,
  pp. 1276--1282.

\bibitem{kowalski_regular_2019}
J.~Kowalski, M.~Mika, J.~Sutowicz, and M.~Szykuła, ``Regular boardgames,''
  \emph{Proc. {AAAI} Conf. Artificial Intelligence}, vol.~33, pp. 1699--1706,
  2019.

\bibitem{bonanno_set-theoretic_1992}
G.~Bonanno, ``Set-theoretic equivalence of extensive-form games,''
  \emph{International Journal of Game Theory}, vol.~20, no.~4, pp. 429--447,
  1992.

\bibitem{chvatil_mage_2011}
V.~Chvátil, \emph{Mage {K}night}.\hskip 1em plus 0.5em minus 0.4em\relax
  {WizKids}, 2011, board game.

\end{thebibliography}

\end{document}